Rotational spectrum of isotopic methyl mercaptan, $^{13}$CH$_3$SH, in the laboratory and towards Sagittarius B2(N2)[1]

Vadim V. Ilyushin[1,2], Olena Zakharenko[3], Frank Lewen[3], Stephan Schlemmer[3], Eugene A. Alekseev[1,2], Mykola Pogrebnyak[1,2], Ronald M. Lees[4], Li-Hong Xu[4†], Arnaud Belloche[5], Karl M. Menten[5], Robin T. Garrod[6], and Holger S. P. Müller[3]

[1]Institute of Radio Astronomy of NASU, Mystetstv 4, 61002 Kharkiv, Ukraine

[2] Quantum Radiophysics Department, V. N. Karazin Kharkiv National University, Svobody Square 4, 61022 Kharkiv, Ukraine.

[3] I. Physikalisches Institut, Universität zu Köln, Zülpicher Str. 77, 50937 Köln, Germany.

[4]Department of Physics, University of New Brunswick, Saint John, NB, Canada.

[5] Max-Planck-Institut für Radioastronomie, Auf dem Hügel 69, 53121 Bonn, Germany.

[6] Departments of Chemistry and Astronomy, The University of Virginia, Charlottesville, VA, USA.

**Corresponding authors:** V. V. Ilyushin (email: ilyushin@rian.kharkov.ua) and H. S. P. Müller (email: hspm@ph1.uni-koeln.de)

[1]This manuscript is part of a special issue to honor Professor Li-Hong Xu.
† Deceased.

**Abstract:** Methyl mercaptan ($CH_3SH$) is a known interstellar molecule with abundances high enough that the detection of some of its minor isotopologues is promising. The present study aims to provide accurate spectroscopic parameters for the $^{13}CH_3SH$ isotopologue to facilitate its identification in the interstellar medium at millimetre and submillimetre wavelengths. Through careful analysis of recent $CH_3SH$ spectra from 49−510 GHz and 1.1−1.5 THz recorded at natural isotopic composition, extensive assignments were possible not only for the ground torsional state of $^{13}CH_3SH$, but also in the first and second excited states. The torsion-rotation spectrum displays complex structure due to the large-amplitude internal rotation of the $^{13}CH_3$ group, similar to the main and other minor isotopic species of methyl mercaptan. The assigned transition frequencies have been fitted to within experimental error with a 52-parameter model employing the RAM36 programme. With predictions based on this fit, $^{13}CH_3SH$ was searched for in spectra from the Atacama Large Millimetre/submillimetre Array (ALMA) towards the Galactic centre source Sgr B2(N2). Several transitions were expected to be observable, but all of them turned out to be severely blendedwith emission from other species, which prevents us from identifying $^{13}CH_3SH$ in this source.

# 1. Introduction

Sulphur-bearing molecules account for more than 20 of the roughly 200 molecules discovered in the interstellar medium or in the circumstellar envelopes of late-type stars [1, 2]. Some pairs of them were proposed to serve as chemical clocks in the evolution of protostars [3, 4, and references therein]. However, the chemistry in star-forming regions has been poorly understood at least until quite recently because the abundances of the observed sulphur-bearing molecules in these sources accounted only for a small fraction of the total sulphur budget whose depletion into dust grains has been poorly constrained [5−7]. Considerable progress has been made over the years, particularly in a recent contribution [8].

Methyl mercaptan, $CH_3SH$, also known as methanethiol, was one of the molecules discovered early in space by radio-astronomical means, initially only tentatively [9], but confirmed subsequently [10]. Both observations were made towards the high-mass star-forming region Sagittarius (Sgr) B2 close to the Galactic centre. The molecule has been found towards several other star-forming regions since then [11−15]. The Atacama Large Millimetre /submillimetre Array (ALMA) was used recently forunbiased molecular line surveys towards Sgr B2(N) [16] and IRAS 16293−2422 [17]. Methyl mercaptan was found at levels at which minor isotopic species could be detected as well.

A large variety of organic molecules in which one $^{12}C$ atom is substituted by $^{13}C$ have been detected in the interstellar medium over the years, many of them recently. They include $^{13}C$ isotopologues of ethanol [16], glycolaldehyde [18] and formic acid, ketene and acetaldehyde [19]. Even methyl cyanide [20] and ethyl cyanide with two $^{13}C$ [21], cyclopropenylidene with $^{13}C$ and D [22] and formaldehyde with $^{13}C$ and two D [23] have been detected. These findings were critically dependent on a knowledge of rest frequencies with high accuracy, generated usually from laboratory measurements and provided in databases such as the Cologne Database for Molecular Spectroscopy, CDMS [24, 25]. Such laboratory data were published only quite recently in several cases, sometimes even together with the astronomical study. We list here the most relevant publications for the appropriate isotopologues of ethanol [26], glycolaldehyde [27], formic acid [28], ketene [29], acetaldehyde [30], methyl cyanide [31], ethylcyanide [21], cyclopropenylidene [32] and formaldehyde [33]. The detection of so many molecules with one or even two $^{13}C$ may be surprising given the terrestrial $^{12}C/^{13}C$ ratio of 89 [34]. However, this ratio is only about 20 to 25 in the Galactic centre region [16, 20, 35], greatly increasing the chances of finding isotopologues with $^{13}C$ in this region. There is a gradient in the Galactic $^{12}C/^{13}C$ ratio which increases to around 68 in the interstellar medium in the Solar neighbourhood and even further in the outskirts of the Milky Way [35−37].

Methyl mercaptan is a heavier homologue of methanol, $CH_3OH$. Initial investigations into the rotation-torsional spectrum of $CH_3SH$ in the microwave and lower millimetre-wave regions [38−41] were devoted to the determination of its structural and internal rotation parameters. Later millimetre-wave studies [42, 43] as well as a work by some of the current authors covering sections of the 1.1−1.8 THz region [44] were carried out to provide accurate transition frequencies for radio-astronomical observations. More recently, we performed new measurements on $CH_3SD$ [45] in the 150−510 GHz region to facilitate astronomical searches for this methyl mercaptan isotopic species. In addition, we recorded new spectra of $CH_3SH$ covering 49−510 GHz to reach higher $J$ and $K$ rotational quantum numbers for astronomical observations [46], and also to analyse the intricate perturbations between higher torsional states and lower-lying small amplitude vibrations that manifest themselves already in lower torsional states, especially $v_t = 2$, via torsional interactions. These spectra together with our earlier recordings were sensitive enough to identify a large body of transition frequencies pertaining to $CH_3^{34}SH$ in natural isotopic composition in the lowest threetorsional states [47].

Some lines were subsequently identified tentatively in our ALMA data at about the expected $^{32}$S/$^{34}$S isotopic ratio, but the number of unblended or slightly blended transitions was too small to report a secure detection of this isotopologue [47].

In the present work, we report the results of searches for rotation-torsional transitions of $^{13}$CH$_3$SH in the spectral recordings of CH$_3$SH in natural isotopic composition. Previously published data on this isotopologue are, to the best of our knowledge, very limited. Only five transition frequencies were reported [38], all pertaining to the $J = 1-0$ $a$-type transition. The A and E symmetry components were reported for the ground and first excited torsional states, and only the A component for the second excited torsional state. We identified transitions for a wide range of $J$ and $K$ extending up to $v_t = 2$ in the 49−510 GHz and for $v_t = 0$ in the 1.1−1.5 THz recordings. These data were modelled in a very similar way to the previous CH$_3$SD, CH$_3$SH and CH$_3$$^{34}$SH analyses [45−47]. In this work we report also on our search for this isotopic species in our ALMA data taken towards Sgr B2(N).

## 2. Experimental details

New measurements in Cologne were carried out covering almost all of the 155−510 GHz frequency range using the Cologne millimetre/sub-millimetre wave spectrometer. The frequency source was a set of VDI multiplier chains (Virginia Diodes, Inc.) driven by an Agilent E8257D synthesizer referenced to a rubidium standard. Schottky diodes were used for detection, with the lock-in amplifier set to 2$f$ mode giving close to the second derivative of the actual lineshape. The 5-m double-pass glass cell was maintained at room temperature at pressures of 2−4 Pa. A CH$_3$SH sample ($\geq$ 98%) from Sigma-Aldrich was used without further purification and the $^{13}$C isotopic species was measured in natural abundance. Additional information on the spectrometer system is available in [48]. Uncertainties of 30 kHz were assigned to stronger, isolated lines; larger uncertainties up to 200 kHz were attributed to weaker lines or lines close to others.

The sources used for the earlier measurements in the 1.1−1.5 THz region were similar to those above, except that higher harmonics had to be generated. A liquid helium cooled InSb bolometer (QMC Instruments Ltd.) was employed as detector. The measurements employed the same sample as above at pressures of around 2 Pa, albeit in a 3-m single-pass glass cell. Further details on the spectrometer system are available elsewhere [44]. Accuracies of 10 kHz can be achieved with this spectrometer [49, 50] but for the weak lines of $^{13}$CH$_3$SH in the present study, we attributed uncertainties of 100 kHz and 200 kHz.

Measurements in the 49−150 GHz range were performed in Kharkiv on the automated spectrometer of the Institute of Radio Astronomy of NASU [51], employing two-step frequency multiplication of a reference synthesizer (385−430 MHz) in two phase-lock-loop stages. The first multiplication step is formed by a klystron that operates in the 3.4−5.2 GHz frequency range and has a narrowband (1 kHz) phase-lock-loop system. Several backward wave oscillatorsare employedat the second multiplication stage to reach frequencies from 49 to 149 GHz. Frequency modulation was also used in Kharkiv, but with the lock-in amplifier in 1$f$ mode giving close to a first-derivative lineshape. Addition of HCl to a 21% water solution of sodium thiomethoxide, CH$_3$SNa, (purchased from Sigma Aldrich) yielded methyl mercaptan, which was used without further purification. The measurements were performed at room temperature and at pressures in the 1−2 Pa range. The $^{13}$C isotopic species of CH$_3$SH was measured in natural abundance, with uncertainties estimated as 10 kHz for relatively strong isolated lines (S/N > 10), 30 kHz for weak lines (2 < S/N ratio < 10) and 100 kHz for very weak lines (S/N < 2).

## 3. Theoretical model

In the present investigation we used the rho-axis-method (RAM) previously applied successfully to the analysis of methyl mercaptan and some of its isotopologues [44−47]. The Hamiltonian is based on the work of Kirtman [52], Lees and Baker [53], and Herbst et al. [54] and has proven to be effective for molecules containing a $C_{3v}$ rotor and $C_s$ frame. Our rho-axis-method for 3 and 6 fold barriers (RAM36) code implements the RAM approach for molecules with the $C_{3v}$ top attached to a molecular frame of either $C_s$ or $C_{2v}$ symmetry and having either a 3- or 6-fold barrier to internal rotation respectively [55, 56].

The RAM36 Hamiltonian is represented as:

$$H = (1/2) \sum_{knpqrst} B_{knpqrst} [J^{2k}J_z^n J_x^p J_y^q p_\alpha^r \cos(3s\alpha)\sin(3t\alpha) + \sin(3t\alpha)\cos(3s\alpha)p_\alpha^r J_y^q J_x^p J_z^n J^{2k}] \quad (1)$$

where the $B_{knpqrst}$ are fitting parameters; $p_\alpha$ is the angular momentum conjugate to the internal rotation angle $\alpha$; $J_x$, $J_y$, $J_z$ are projections on the $x,y,z$ axes of the total angular momentum $J$. The expression is written in a slightly symbolic form, in the sense that terms generated from the input with $t = 0$ are obtained by omitting the corresponding sine functions, rather than by simply setting $t = 0$ in them. In the case of a $C_{3v}$ top and $C_s$ frame applicable to methyl mercaptan, allowed terms in the torsion-rotation Hamiltonian must be totally symmetric in the group $G_6$ (and also must be Hermitian and invariant to the time reversal operation). Further details on the RAM36 code can be found in the literature [55,56].

## 4. Assignments and fitting

The new data were assigned starting from the Cologne measurements in the 155−510 GHz frequency range and searching for series of intense $a$-type $R$-branch transitions with predictions based on rotational constants obtained from low level quantum chemical calculations. Assuming that the $^{13}$C isotopic substitution does not appreciably alter either the molecular structure or the force field of the molecule, the quartic centrifugal distortion parameters and main internal rotation parameters (internal rotation constant $F$, coupling parameter $\rho$, internal rotation barrier $V_3$) were fixed at the corresponding values of the $^{12}$CH$_3$SH parent molecule [46]. Fig. 1 illustrates the initial search of $^{13}$CH$_3$SH transitions in the 170−180 GHz range showing the relative intensities of the series of $^aR6$ rotational transitions in the main, $^{34}$S and $^{13}$C isotopologues of methyl mercaptan. The similarity with the spectral patterns of the $^{34}$S and the main isotopic species allowed rather straightforward assignments of ground state lines of $^{13}$CH$_3$SH and obtaining first fits. Further analysis proceeded as usual by gradually adding newly assigned transitions to the dataset in iterative cycles of improving the parameter set. In these cycles the data belonging to the first and second excited torsional states of $^{13}$CH$_3$SH were also added to the analysis. The Kharkiv measurements in the 49−149 GHz range were analyzed subsequently, and then the terahertz records (1.1−1.5 THz) were assigned at the final stage.

In total, 2702 rotational transitions were measured for the ground, first and second excited torsional states of $^{13}$CH$_3$SH which correspond to 2319 different line frequencies because of blending. Table 1 gives an overview of the dataset and the quality of our "best fit" in the present study. The highest $J$ and $K_a$ quantum numbers in the fit obviously decrease with increasing torsional excitation. The full dataset was fitted using 52 parameters with an overall weighted root mean square (rms) deviation of 0.85. The input and output files of the final fit are available as supplementary material with this article. The 52 parameters from the global fit of the $^{13}$CH$_3$SH spectrum are listed in Table 2. Figs. 2 and 3 visualise the agreement that is achieved with our current model for the most intense $R$-type transitions around 121.9 GHz

($^aR$4 series) and 316.9 GHz ($^aR$12 series). It is seen that the positions and relative intensities of the $^{13}$CH$_3$SH lines are well predicted by our current model.

In Table 3 we compare the molecular parameters of all isotopologues studied by us thus far with the RAM approach [45−47]. Differences in the datasets led to slightly different sets of high order torsion-rotational parameters in all four fits. The comparison in Table 3 is consequently limited to parameters up to fourth order. Note that some parameter and operator expressions in Table 3 have changed in comparison with Table 2. This is caused by the fact that the general Hamiltonian form (1) encoded in the RAM36 programme does not allow any modification of the coefficient in front of the expression. Thus all coefficients historically adopted for a number of terms (such as the minus sign in front of the quartic centrifugal distortion terms) are absorbed in the parameter values. For the purpose of the current comparison, we have recalculated all differing parameters to a more traditional form of the work [44], which may be considered as a starting point of our methyl mercaptan studies.

As we can see from Table 3, there are no big changes in the main internal rotational parameters $V_3$, $\rho$, and $F$ between the main, $^{13}$C, and $^{34}$S isotopologues compared to the H/D substitution. The $^{12}$C/$^{13}$C and $^{32}$S/$^{34}$S substitutions affect mainly $B$ and $C$ among the rotational constants, whereas H/D substitution significantly affects also the $A$ rotational constant and $D_{ab}$ (the change in sign of $D_{ab}$ was discussed in detail in ref. [45]). Thus it is not surprising that the $D_K$ parameter of CH$_3$SD is the most deviating among the quartic centrifugal distortion parameters, which in general agree rather well especially for the main, $^{13}$C and $^{34}$S isotopologues. Substantial differences are observed for the $V_6$ term in the torsional potential expansion between the main and all other isotopologues. This is a consequence of differences in the datasets, because for the main isotopologue we were able to reach much higher energy levels than in the case of the $^{13}$C, $^{34}$S and D isotopologues. The higher we go up in energy the stronger the influence of intervibrational interactions with non-torsional low-lying vibrational states in methyl mercaptan (C−S stretch and ∠CSH bend). In the current model, which does not explicitly include interactions with low-lying vibrational states, this influence is partly absorbed by the expansion coefficients of the torsional potential function distorting their 'true' values (this phenomenon is discussed in more details in Ref. [46]). In the case of $^{13}$C, $^{34}$S and D isotopologues these distortions are much smaller since we were not able to reach high enough energies with our assignments. We point out that our first attempts to include in the Hamiltonian model explicit interactions with low lying vibrational states [58] lead to significant reductions in the $V_6$ value, making it close to $V_6$ value of the $^{34}$S isotopologue.

If we consider in turn fourth order torsion-rotation distortion parameters we see a very good agreement between main and $^{13}$C, $^{34}$S isotopologues for $V_{3J}$, $V_{3K}$, $V_{3bc}$, $V_{3ab}$, $F_J$, $F_K$, $D_{3ac}$, $\rho_J$, and $\rho_K$ parameters and main torsional parameters (especially $F$ and $\rho$). The D isotopologue shows more discrepancies in these parameters than the three other isotopologues, as can be expected because of the changes in the rotational parameters. The remaining fourth order torsion-rotation distortion parameters, namely $F_{ab}$, $F_{bc}$, $D_{3bc}$, $\rho_{ab}$, $\rho_{bc}$, $D_{abJ}$, and $D_{abK}$, display more disagreement in their values, and only $D_{3bc}$ is present in all parameter sets (see Table 3). So, it is quite possible that the observed disagreement in parameter values of different isotopologues is caused in part by differences in the parameter sets and thus further insights into the intercomparison of the RAM Hamiltonians of different methyl mercaptan isotopologues may be obtained only after unification of the parameter sets of different isotopologues (which is beyond the scope of the current study and may not be possible with the present datasets).

## 5. Astronomical observation

In our search for the $^{13}$C isotopologue of methyl mercaptan we employ the Exploring Molecular Complexity with ALMA (EMoCA) imaging spectral line survey performed towards the high-mass star-forming region Sgr B2(N) using ALMA. The description of the observations, data reduction as well as the method used to identify the detected lines and derive column densities may be found in ref. [20]. In brief, the frequency range from 84.1 GHz to 114.4 GHz was recorded with five frequency tunings at a spectral resolution of 488.3 kHz (1.7 to 1.3 km s$^{-1}$). The survey has a median angular resolution of 1.″6. The phase centre was set at $(\alpha,\delta)_{J2000}$ = ($17^h47^m19.^s87$, $-28°22'16.″0$). We focus on the peak position of the hot molecular core Sgr B2(N2) at $(\alpha,\delta)_{J2000}$ = ($17^h47^m19.^s86$, $-28°22'13.″4$) because the narrower lines and the absence of line wings in Sgr B2(N2) are favourable for detecting molecular species with weak emission lines. Important findings towards this core include the first detection of a branched alkyl compound, namely isopropyl cyanide [59], the detection of ethyl cyanide isotopomers with two $^{13}$C [21], an account on deuteration in Sgr B2(N2) [20] and an account on alkanethiols and alkanols in Sgr B2(N2) [16].

## 6. Astronomical results

The best-fit radiative-transfer model obtained for the $^{32}$S main isotopic species for Sgr B2(N2) [16] was used as a starting point in our search for rotational lines of the $^{13}$C isotopologue of methyl mercaptan in the EMoCA spectrum of this source. We assumed the same temperature (180 K), size of the emitting region (1.″4), linewidth (5.4 km s$^{-1}$), and centroid velocity (73.5 km s$^{-1}$) as those derived for the main isotopologue. We assumed a $^{12}$C/$^{13}$C ratio of 23 based on findings that this ratio is about 20 to 25 in the Galactic centre [16, 20, 35-37]. With this assumption we get a column density of $1.5\times10^{16}$ cm$^{-2}$ for $^{13}$CH$_3$SH (from $3.4\times10^{17}$ cm$^{-2}$ obtained for the main isotopologue). These parameters were taken to model a synthetic spectrum of $^{13}$CH$_3$SH using Weeds [60] under the assumption of the local thermodynamic equilibrium approximation. The result is compared to the observed EMoCA spectrum in Fig. 4. The synthetic spectrum of $^{13}$CH$_3$SH contains some emission features around three times the average noise level of Sgr B2(N2), however all of the strongest transitions predicted for $^{13}$CH$_3$SH are, regrettably, blended with emission from other molecules, preventing their attribution to this isotopologue. Therefore, this species cannot be unambiguously identified in the EMoCA spectrum.

## 7. Conclusions

A new investigation into the rotational spectrum of the $^{13}$C isotopologue of methyl mercaptan was performed in the frequency ranges 49−510 GHz and 1.1−1.5 THz in order to derive accurate rest frequencies for astronomical searches. The rotational transitions of the ground, first and second excited torsional states with $J$ up to 55 were fitted using a set of 52 RAM Hamiltonian parameters with an overall weighted rms deviation of 0.85. Reliable rest frequency predictions were calculated for astrophysical use up to 1.5 THz based on the present results. These are available as supplementary material to this article and will also be available in the catalogue section of the CDMS [24, 25]. $^{13}$CH$_3$SH was searched for in the EMoCA spectrum of Sgr B2(N2) between 84 GHz and 114 GHz, however blends with emission lines of other molecules impede its detection in this frequency range for this source. A future detection requires observations at higher angular resolution, at other frequencies or towards other sources.

**Acknowledgements**

The present investigation benefitted from funding by the Deutsche Forschungsgemeinschaft (DFG) within the scope of the collaborative research grant SFB 956 (project ID 184018867), sub-project B3. OZ received support from the DFG via the Gerätezentrum "Cologne Center for Terahertz Spectroscopy" (project ID SCHL 341/15-1). The studies in Kharkiv were supportedby the Volkswagen foundationwith assistance fromthe Science and Technology Center in the Ukraine (STCU partner project P686). RML and LHX are grateful for assistance by the Natural Sciences and Engineering Research Council of Canada.

Figure captions.

Fig. 1. Part of the spectrum showing the $^aR6$ rotational transitions for the main, $^{34}$S and $^{13}$C isotopologues of methyl mercaptan. [Colour online.]

Fig. 2. Comparison of the observed (upper) and the calculated (lower) spectra of $^{13}$CH$_3$SH near 121.9 GHz recorded in first derivative detection mode of the Kharkiv spectrometer. At the top of this figure the symmetry (A or E) and $K_a$ assignments of the $J = 5 \leftarrow 4$ series of transitions dominating this part of the spectrum are given.

Fig. 3. Comparison of the observed (upper) and the calculated (lower) spectra of $^{13}$CH$_3$SH near 316.9 GHz recorded in second derivative detection mode of the Cologne spectrometer. At the top of this figure the symmetry (A or E) and $K_a$ assignments of the $J = 13 \leftarrow 12$ series of transitions dominating this part of the spectrum are given.

Fig. 4. Parts of the EMoCA spectrum of Sgr B2(N2) that cover frequencies of transitions of $^{13}$CH$_3$SH that are expected to exceed two times the average noise level. The local thermodynamic equilibrium synthetic spectrum of $^{13}$CH$_3$SH is displayed in red and overlaid on the observed spectrum shown in black. The green synthetic spectrum contains the contributions of all molecules identified in our survey so far, but not $^{13}$CH$_3$SH. The *y*-axis is labelled in brightness temperature units. The dotted line indicates the $3\sigma$ noise level. For interpretation of the references to colour in this figure legend, the reader is referred to the web version of this article. [Colour online.]

Fig.1

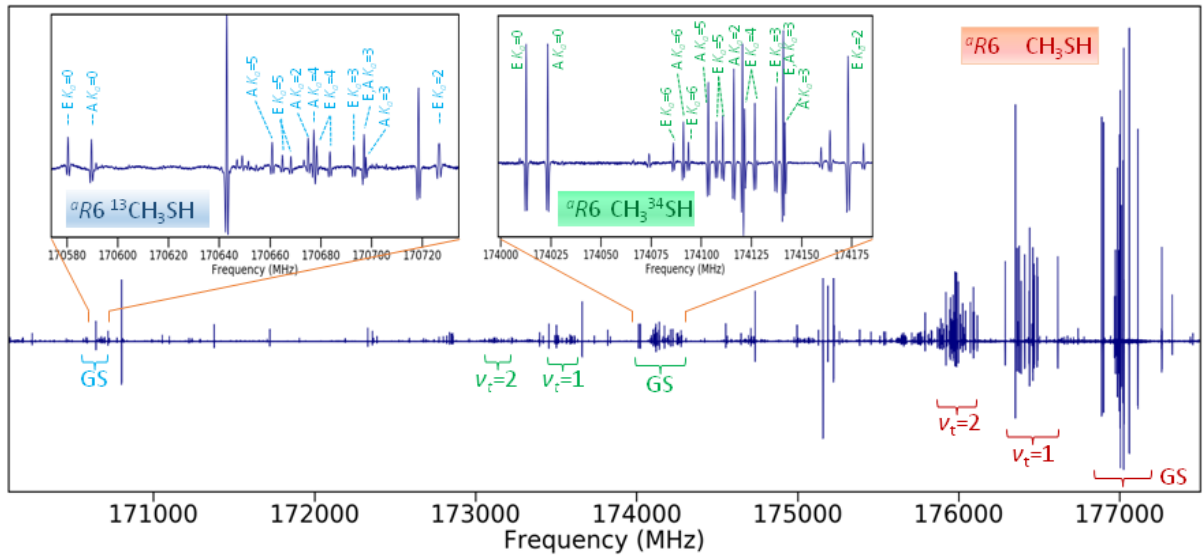

Fig. 2

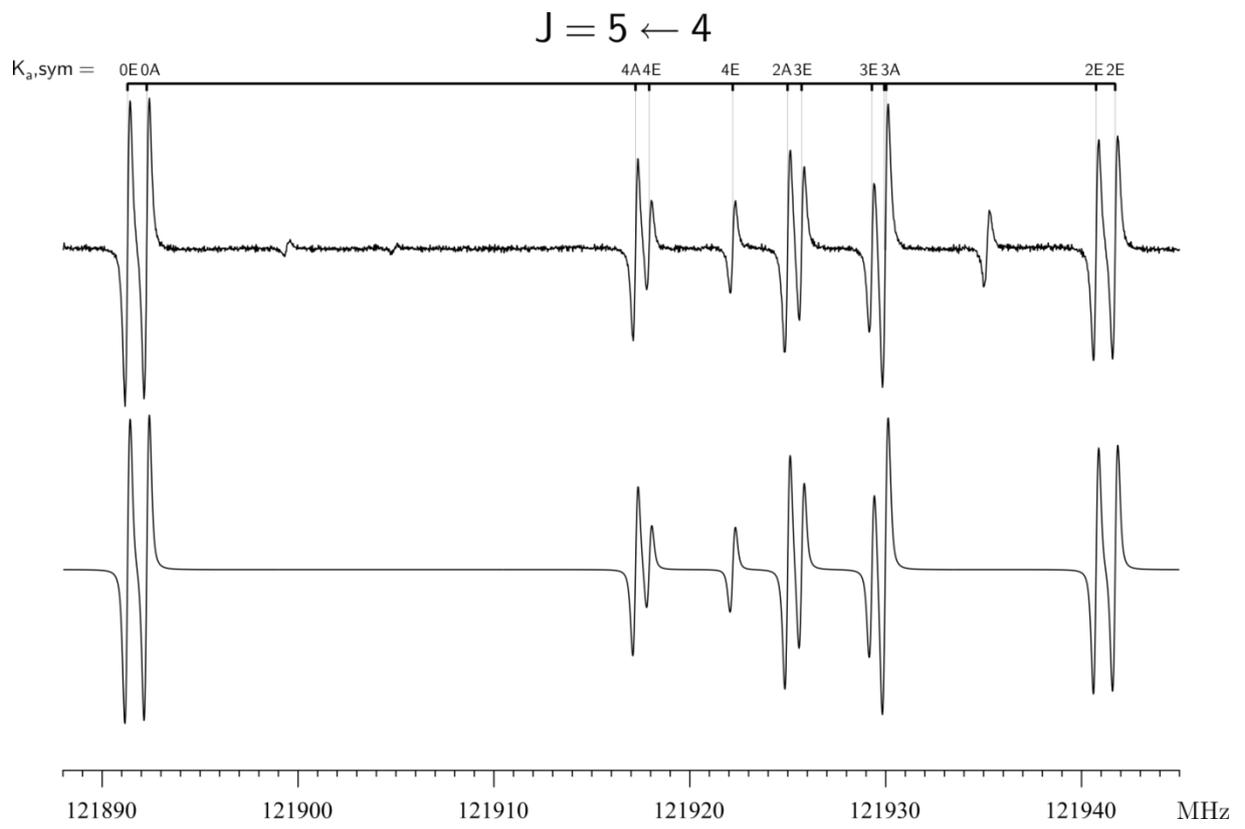

Fig. 3.

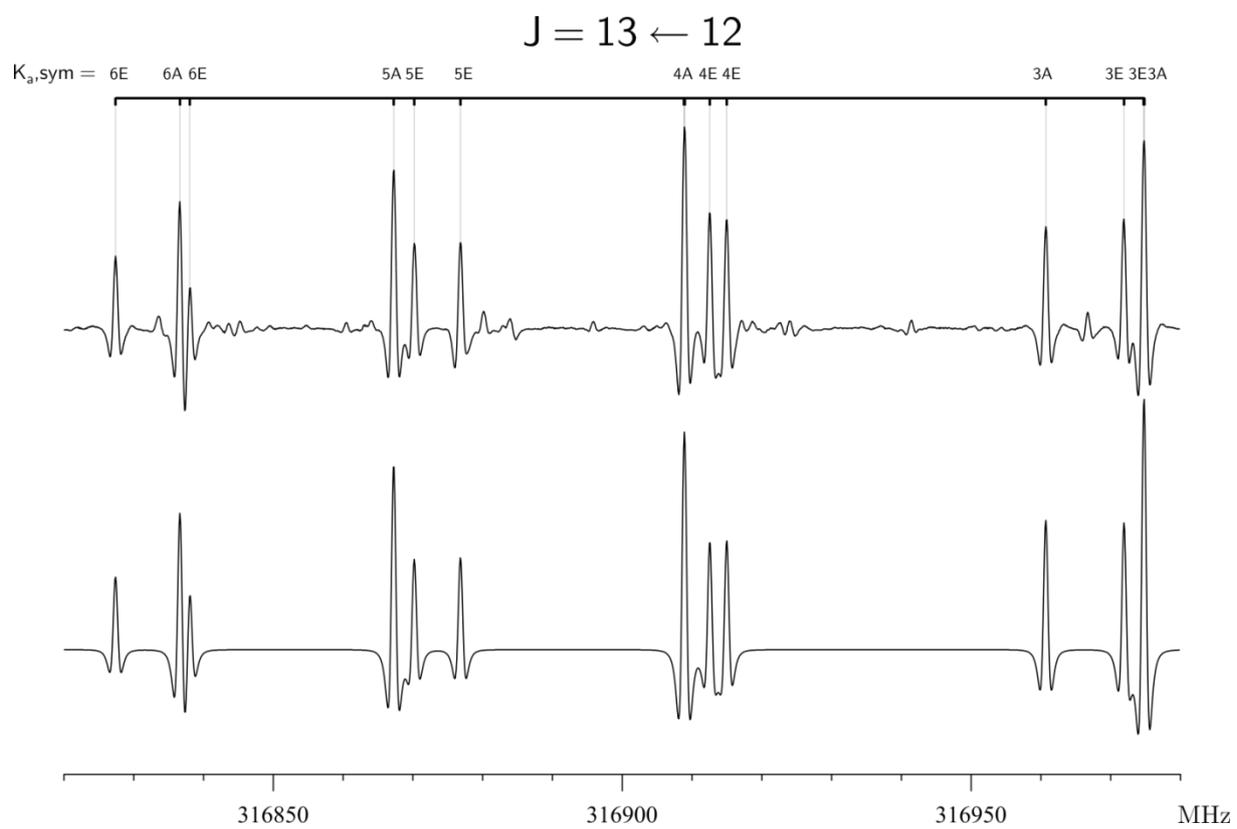

Fig. 4.

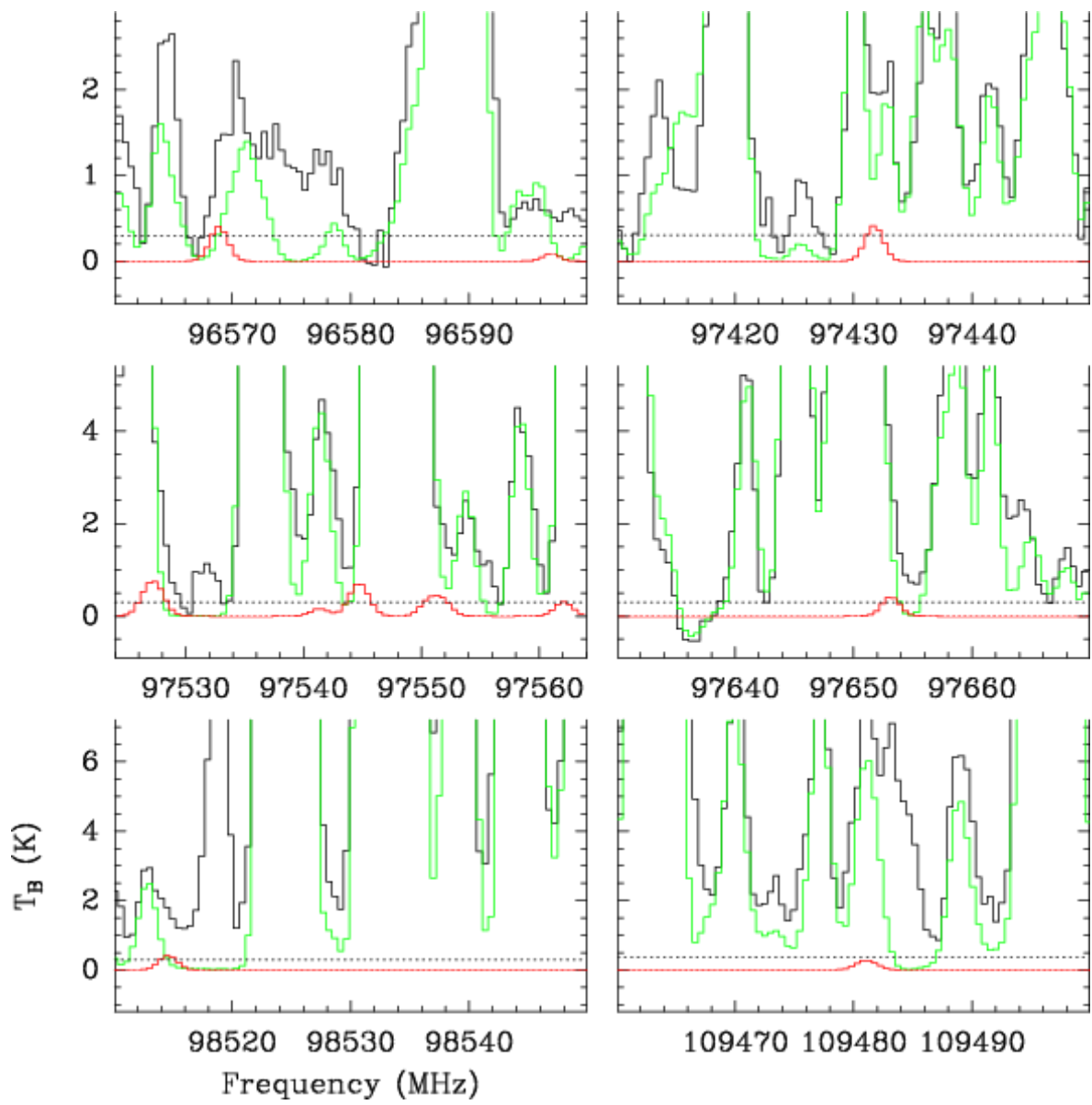

Table 1

Synopsis of the data set and the fit quality.

| By measurement uncertainty | | | By torsional state | | | |
|---|---|---|---|---|---|---|
| Unc.[a] | #[b] | rms[c] | Category[d] | #[b] | wrms[e] | $J, K_a$[f] |
| 0.010 MHz | 129 | 0.0073 MHz | $v_t = 0 - 0$ | 1496 | 0.81 | 55, 15 |
| 0.030 MHz | 888 | 0.0270 MHz | $v_t = 1 - 1$ | 718 | 0.90 | 25, 13 |
| 0.050 MHz | 466 | 0.0465 MHz | $v_t = 2 - 2$ | 488 | 0.97 | 21, 8 |
| 0.100 MHz | 431 | 0.0787 MHz | A-type | 1397 | 0.87 | |
| 0.200 MHz | 405 | 0.1487 MHz | E-type | 1305 | 0.86 | |

[a]Estimated measurement uncertainties for each data group.

[b]Number of lines (left part) or transitions (right part) of each category in the least-squares fit. The 2702 transitions correspond to 2319 different line frequencies in the fit because of blending, caused for example by unresolved asymmetry splitting.

[c]Root-mean-square (rms) deviation of corresponding data group.

[d]The transitions are grouped by torsional quantum number $v_t$ followed by symmetry species.

[e]Weighted root-mean-square (wrms) deviation of the corresponding data group.

[f]The maximum value of $J$ and $K_a$ for transitions in each torsional state in the final fit.

Table 2
Fitted parameters of the RAM Hamiltonian for $^{13}$CH$_3$SH molecule.

| Operator[a] | $n_{tr}$[b] | Parameter[c] | Value[d] |
|---|---|---|---|
| $p_\alpha^2$ | $2_{20}$ | $F$ | 15.038061(60) |
| $(1-\cos(3\alpha))$ | $2_{20}$ | $(1/2)V_3$ | 220.8666(14) |
| $p_\alpha P_a$ | $2_{11}$ | $\rho$ | 0.65182799(21) |
| $P_a^2$ | $2_{02}$ | $A_{RAM}$ | 3.426214(42) |
| $P_b^2$ | $2_{02}$ | $B_{RAM}$ | 0.416148967(78) |
| $P_c^2$ | $2_{02}$ | $C_{RAM}$ | 0.39868399(12) |
| $(\frac{1}{2})\{P_a,P_b\}$ | $2_{02}$ | $2D_{ab}$ | $-0.0145488(45)$ |
| $(1-\cos(6\alpha))$ | $4_{40}$ | $(1/2)V_6$ | $-0.3409(38)$ |
| $p_\alpha^4$ | $4_{40}$ | $F_m$ | $-0.11608(41)\times10^{-2}$ |
| $p_\alpha^3 P_a$ | $4_{31}$ | $\rho_m$ | $-0.3649(11)\times10^{-2}$ |
| $P^2(1-\cos(3\alpha))$ | $4_{22}$ | $V_{3J}$ | $-0.1994380(13)\times10^{-2}$ |
| $P_a^2(1-\cos(3\alpha))$ | $4_{22}$ | $V_{3K}$ | $0.7870(22)\times10^{-2}$ |
| $(P_b^2-P_c^2)(1-\cos(3\alpha))$ | $4_{22}$ | $V_{3bc}$ | $-0.84652(74)\times10^{-4}$ |
| $(\frac{1}{2})\{P_a,P_b\}(1-\cos(3\alpha))$ | $4_{22}$ | $V_{3ab}$ | $0.119786(26)\times10^{-1}$ |
| $p_\alpha^2 P^2$ | $4_{22}$ | $F_J$ | $-0.288389(14)\times10^{-4}$ |
| $p_\alpha^2 P_a^2$ | $4_{22}$ | $F_K$ | $-0.4870(12)\times10^{-2}$ |
| $p_\alpha^2(P_b^2-P_c^2)$ | $4_{22}$ | $F_{bc}$ | $0.22284(52)\times10^{-4}$ |
| $(\frac{1}{2})\{P_a,P_c\}\sin(3\alpha)$ | $4_{22}$ | $D_{3ac}$ | $0.144880(93)\times10^{-1}$ |
| $(\frac{1}{2})\{P_b,P_c\}\sin(3\alpha)$ | $4_{22}$ | $D_{3bc}$ | $0.15673(18)\times10^{-2}$ |
| $p_\alpha P_a P^2$ | $4_{13}$ | $\rho_J$ | $-0.395418(17)\times10^{-4}$ |
| $p_\alpha P_a^3$ | $4_{13}$ | $\rho_K$ | $-0.29850(56)\times10^{-2}$ |
| $P^4$ | $4_{04}$ | $-\Delta_J$ | $-0.506073(12)\times10^{-6}$ |
| $P^2 P_a^2$ | $4_{04}$ | $-\Delta_{JK}$ | $-0.1659781(67)\times10^{-4}$ |
| $P_a^4$ | $4_{04}$ | $-\Delta_K$ | $-0.7017(11)\times10^{-3}$ |
| $P^2(P_b^2-P_c^2)$ | $4_{04}$ | $-2\delta_J$ | $-0.40483(17)\times10^{-7}$ |
| $(\frac{1}{2})\{P_a^2,(P_b^2-P_c^2)\}$ | $4_{04}$ | $-2\delta_K$ | $-0.24156(47)\times10^{-4}$ |
| $(1-\cos(9\alpha))$ | $6_{60}$ | $(1/2)V_9$ | 0.2274(52) |
| $P^2(1-\cos(6\alpha))$ | $6_{42}$ | $V_{6J}$ | $-0.14583(33)\times10^{-4}$ |
| $P_a^2(1-\cos(6\alpha))$ | $6_{42}$ | $V_{6K}$ | $-0.1780(50)\times10^{-2}$ |
| $(\frac{1}{2})\{P_a,P_b\}(1-\cos(6\alpha))$ | $6_{42}$ | $V_{6ab}$ | $0.1168(20)\times10^{-3}$ |
| $p_\alpha^4 P^2$ | $6_{42}$ | $F_{mJ}$ | $0.1269(24)\times10^{-8}$ |
| $(\frac{1}{2})\{P_b,P_c\}\sin(6\alpha)$ | $6_{42}$ | $D_{6bc}$ | $0.12542(49)\times10^{-3}$ |
| $(\frac{1}{2})\{P_a,P_c\}\sin(6\alpha)$ | $6_{42}$ | $D_{6ac}$ | $0.3287(67)\times10^{-3}$ |
| $(\frac{1}{2})\{P_b,P_c,p_\alpha^2,\sin(3\alpha)\}$ | $6_{42}$ | $D_{3bcm}$ | $0.3358(11)\times10^{-4}$ |
| $p_\alpha^3 P_a P^2$ | $6_{33}$ | $\rho_{mJ}$ | $0.1089(20)\times10^{-8}$ |
| $P^4(1-\cos(3\alpha))$ | $6_{24}$ | $V_{3JJ}$ | $0.4410(17)\times10^{-8}$ |
| $P^2 P_a^2(1-\cos(3\alpha))$ | $6_{24}$ | $V_{3JK}$ | $-0.24338(54)\times10^{-6}$ |
| $(\frac{1}{2})\{P_a^2,(P_b^2-P_c^2)\}(1-\cos(3\alpha))$ | $6_{24}$ | $V_{3bcK}$ | $0.3118(12)\times10^{-5}$ |
| $(\frac{1}{2})\{P_a,P_c\}P^2\sin(3\alpha)$ | $6_{24}$ | $D_{3acJ}$ | $-0.2202(48)\times10^{-6}$ |
| $(\frac{1}{2})\{P_b,P_c\}P^2\sin(3\alpha)$ | $6_{24}$ | $D_{3bcJ}$ | $-0.1692(10)\times10^{-7}$ |
| $(\frac{1}{2})\{P_a^2,P_b,P_c\}\sin(3\alpha)$ | $6_{24}$ | $D_{3bcK}$ | $-0.12683(50)\times10^{-4}$ |
| $(\frac{1}{2})(\{P_b^3,P_c\}-\{P_b,P_c^3\})\sin(3\alpha)$ | $6_{24}$ | $D_{3bcbc}$ | $-0.586(10)\times10^{-8}$ |
| (½){$P_a,P_b^3$}cos(3α) | $6_{24}$ | $V_{3ab3}$ | 0.2236(42)×10$^{-6}$ |
| (½){$P_c^2,P_b^2$}$p_α^2$ | $6_{24}$ | $F_{b2c2}$ | 0.737(13)×10$^{-9}$ |
| (½){$P_a,P_b^2,P_c^2$}$p_α$ | $6_{15}$ | $\rho_{b2c2}$ | 0.946(15)×10$^{-9}$ |
| $P^6$ | $6_{06}$ | $\Phi_J$ | −0.2681(15)×10$^{-12}$ |
| $P^4 P_a^2$ | $6_{06}$ | $\Phi_{JK}$ | 0.7230(73)×10$^{-10}$ |
| $P_a^6$ | $6_{06}$ | $\Phi_K$ | 0.555(22)×10$^{-9}$ |
| $P^2(1-\cos(9α))$ | $8_{62}$ | $V_{9J}$ | −0.1950(69)×10$^{-5}$ |
| $P_a^2(1-\cos(9α))$ | $8_{62}$ | $V_{9K}$ | 0.2955(76)×10$^{-2}$ |
| (½){$P_b,P_c$}sin(9α) | $8_{62}$ | $D_{9bc}$ | 0.386(15)×10$^{-4}$ |
| (½){$P_b,P_c$}$P^2$sin(6α) | $8_{44}$ | $D_{6bcJ}$ | 0.264(15)×10$^{-8}$ |

[a] {A,B,C,D}=ABCD + DCBA. {A,B,C}=ABC + CBA. {A,B}=AB + BA. The product of the operator in the first column of a given row and the parameter in the third column of that row gives the term actually used in the torsion-rotation Hamiltonian of the program, except for $F$, $\rho$ and $A_{RAM}$, which occur in the Hamiltonian in the form $F(p_α+\rho P_a)^2 + A_{RAM} P_a^2$.

[b] $n = t + r$, where $n$ is the total order of the operator, $t$ is the order of the torsional part and $r$ is the order of the rotational part, respectively.

[c] Parameter nomenclature based on the subscript procedures of [57].

[d] All values are in cm$^{-1}$, except $\rho$ which is unitless. Statistical uncertainties are shown as one standard uncertainty in the units of the last two digits.

Table 3. Comparison of the low-order parameters of $CH_3SH$, $^{13}CH_3SH$, $CH_3^{34}SH$, and $CH_3SD$.

| Operator[a] | $n_{tr}$[b] | Parameter[c] | $CH_3SH$[d] | $^{13}CH_3SH$[d] | $CH_3^{34}SH$[d] | $CH_3SD$[d] |
|---|---|---|---|---|---|---|
| $p_\alpha^2$ | $2_{20}$ | $F$ | 15.04062399(54) | 15.038061(60) | 15.018101(11) | 10.3520639(31) |
| $(\frac{1}{2})(1-\cos(3\alpha))$ | $2_{20}$ | $V_3$ | 441.69136(24) | 441.7332(27) | 441.56378(72) | 435.42500(24) |
| $p_\alpha P_a$ | $2_{11}$ | $\rho$ | 0.6518557764(11) | 0.65182799(21) | 0.651352093(16) | 0.493517098(11) |
| $P_a^2$ | $2_{02}$ | $A_{RAM}$ | 3.4279249(17) | 3.426214(42) | 3.4254266(18) | 2.59513758(20) |
| $P_b^2$ | $2_{02}$ | $B_{RAM}$ | 0.4320294(32) | 0.416148967(78) | 0.424734752(77) | 0.42517153(13) |
| $P_c^2$ | $2_{02}$ | $C_{RAM}$ | 0.4132203(21) | 0.39868399(12) | 0.406539925(82) | 0.39176839(13) |
| $\{P_a,P_b\}$ | $2_{02}$ | $D_{ab}$ | −0.00737202(42) | −0.0072744(23) | −0.00775864(85) | 0.0053655(12) |
| $(\frac{1}{2})(1-\cos(6\alpha))$ | $4_{40}$ | $V_6$ | −1.9212(12) | −0.6818(77) | −0.56068(40) | −0.86918(20) |
| $p_\alpha^4$ | $4_{40}$ | $F_m$ | −0.1121789(22)×10$^{-2}$ | −0.11608(41)×10$^{-2}$ | −0.11330(13)×10$^{-2}$ | −0.38535(20)×10$^{-3}$ |
| $p_\alpha^3 P_a$ | $4_{31}$ | $\rho_m$ | −0.3554280(60)×10$^{-2}$ | −0.3649(11)×10$^{-2}$ | −0.35743(34)×10$^{-2}$ | −0.93969(42)×10$^{-3}$ |
| $P^2(1-\cos(3\alpha))$ | $4_{22}$ | $V_{3J}$ | −0.2062768(53)×10$^{-2}$ | −0.1994380(13)×10$^{-2}$ | −0.20311808(74)×10$^{-2}$ | −0.19405784(64)×10$^{-2}$ |
| $P_a^2(1-\cos(3\alpha))$ | $4_{22}$ | $V_{3K}$ | 0.7277626(39)×10$^{-2}$ | 0.7870(22)×10$^{-2}$ | 0.71456(12)×10$^{-2}$ | 0.685147(15)×10$^{-2}$ |
| $(P_b^2-P_c^2)(1-\cos(3\alpha))$ | $4_{22}$ | $V_{3bc}$ | −0.81043(38)×10$^{-4}$ | −0.84652(74)×10$^{-4}$ | −0.83920(12)×10$^{-4}$ | −0.15679(13)×10$^{-3}$ |
| $\{P_a,P_b\}(1-\cos(3\alpha))$ | $4_{22}$ | $V_{3ab}$ | 0.614197(19)×10$^{-2}$ | 0.59893(13)×10$^{-2}$ | 0.603843(45)×10$^{-2}$ | 0.443895(41)×10$^{-2}$ |
| $p_\alpha^2 P^2$ | $4_{22}$ | $F_J$ | −0.3094800(28)×10$^{-4}$ | −0.288389(14)×10$^{-4}$ | −0.3000547(70)×10$^{-4}$ | −0.2823067(53)×10$^{-4}$ |
| $p_\alpha^2 P_a^2$ | $4_{22}$ | $F_K$ | −0.4789751(62)×10$^{-2}$ | −0.4870(12)×10$^{-2}$ | −0.47975(33)×10$^{-2}$ | −0.123322(32)×10$^{-2}$ |
| $\{P_a,P_b\}p_\alpha^2$ | $4_{22}$ | $F_{ab}$ | 0.10746(45)×10$^{-4}$ | − | − | 0.7807(70)×10$^{-4}$ |
| $2p_\alpha^2(P_b^2-P_c^2)$ | $4_{22}$ | $F_{bc}$ | −0.32942(41)×10$^{-4}$ | 0.11142(26)×10$^{-4}$ | − | 0.102769(43)×10$^{-4}$ |
| $\{P_a,P_c\}\sin(3\alpha)$ | $4_{22}$ | $D_{3ac}$ | 0.077299(15)×10$^{-1}$ | 0.072440(47)×10$^{-1}$ | 0.072366(16)×10$^{-1}$ | 0.11336(45)×10$^{-1}$ |
| $\{P_b,P_c\}\sin(3\alpha)$ | $4_{22}$ | $D_{3bc}$ | −0.6418(14)×10$^{-3}$ | 0.78365(90)×10$^{-3}$ | 0.450676(38)×10$^{-3}$ | 0.140496(27)×10$^{-2}$ |
| $p_\alpha P_a P^2$ | $4_{13}$ | $\rho_J$ | −0.4255507(38)×10$^{-4}$ | −0.395418(17)×10$^{-4}$ | −0.413515(11)×10$^{-4}$ | −0.369536(28)×10$^{-4}$ |
| $p_\alpha P_a^3$ | $4_{13}$ | $\rho_K$ | −0.2958211(29)×10$^{-2}$ | −0.29850(56)×10$^{-2}$ | −0.29544(14)×10$^{-2}$ | −0.75792(11)×10$^{-3}$ |
| $\{P_a^2,P_b\}p_\alpha$ | $4_{13}$ | $\rho_{ab}$ | 0.10025(43)×10$^{-4}$ | − | − | 0.1017(10)×10$^{-3}$ |
| $\{P_a,(P_b^2-P_c^2)\}p_\alpha$ | $4_{13}$ | $\rho_{bc}$ | −0.42674(40)×10$^{-4}$ | − | −0.09705(39)×10$^{-4}$ | − |
| $-P^4$ | $4_{04}$ | $\Delta_J$ | 0.5393458(89)×10$^{-6}$ | 0.506073(12)×10$^{-6}$ | 0.523063(13)×10$^{-6}$ | 0.4876778(87)×10$^{-6}$ |
| $-P^2 P_a^2$ | $4_{04}$ | $\Delta_{JK}$ | 0.1784933(23)×10$^{-4}$ | 0.1659781(67)×10$^{-4}$ | 0.1737732(50)×10$^{-4}$ | 0.143777(64)×10$^{-4}$ |
| $-P_a^4$ | $4_{04}$ | $\Delta_K$ | 0.6990318(56)×10$^{-3}$ | 0.7017(11)×10$^{-3}$ | 0.69678(23)×10$^{-3}$ | 0.178662(17)×10$^{-3}$ |
| $-2P^2(P_b^2-P_c^2)$ | $4_{04}$ | $\delta_J$ | 0.2281975(75)×10$^{-7}$ | 0.202416(86)×10$^{-7}$ | 0.217630(24)×10$^{-7}$ | 0.384636(10)×10$^{-7}$ |
| $-\{P_a^2,(P_b^2-P_c^2)\}$ | $4_{04}$ | $\delta_K$ | 0.109149(20)×10$^{-4}$ | 0.12078(23)×10$^{-4}$ | 0.09566(37)×10$^{-4}$ | 0.090811(35)×10$^{-4}$ |
| $P^2\{P_a,P_b\}$ | $4_{04}$ | $D_{abJ}$ | − | − | 0.10033(41)×10$^{-6}$ | − |
| $\{P_a^3,P_b\}$ | $4_{04}$ | $D_{abK}$ | − | − | − | 0.2750(34)×10$^{-4}$ |

[a] $\{A,B\}=AB + BA$. The product of the operator in the first column of a given row and the parameter in the third column of that row gives the term actually used in the torsion-rotation Hamiltonian of the program, except for $F$, $\rho$ and $A_{RAM}$, which occur in the Hamiltonian in the form $F(p_\alpha+\rho P_a)^2+A_{RAM}P_a^2$.

[b] $n = t + r$, where $n$ is the total order of the operator, $t$ is the order of the torsional part and $r$ is the order of the rotational part, respectively.

[c] Parameter nomenclature based on the subscript procedures of [57].

[d] All values are in cm$^{-1}$, except $\rho$ which is unitless. Statistical uncertainties are shown as one standard uncertainty in the units of the last two digits.